\documentclass[conference]{IEEEtran}
\usepackage{cite}

\ifCLASSINFOpdf
  \usepackage[pdftex]{graphicx}
  \graphicspath{{/}{../pdf/}{../jpeg/}}
  \DeclareGraphicsExtensions{.pdf,.jpeg,.png,.eps}
\else
  \usepackage[dvips]{graphicx}
  \graphicspath{{/}{../pdf/}{../jpeg/}}
  \DeclareGraphicsExtensions{.eps}
\fi
\usepackage{amsmath}
\usepackage{algorithmic}

\usepackage{array}

\ifCLASSOPTIONcompsoc
 \usepackage[caption=false,font=normalsize,labelfont=sf,textfont=sf]{subfig}
\else
 \usepackage[caption=false,font=footnotesize]{subfig}
\fi
\usepackage{dblfloatfix}

\usepackage{url}

\usepackage{multirow}

\usepackage{color}
\usepackage{cite}

\begin{document}

%
\title{Maximum Value Matters: Finding Hot Topics \\in Scholarly Fields}


\author{\IEEEauthorblockN{Jinghao Zhao, Hao Wu, Fengyu Deng, Wentian Bao, Wencheng Tang, Luoyi Fu, Xinbing Wang}
\IEEEauthorblockA{
Shanghai Jiao Tong University\\
\{zhaojinghao, albert-wuhao, dengfengyu, papermoon, cowentwc, yiluofu, xwang8\}@sjtu.edu.cn}}



%


\IEEEoverridecommandlockouts
\vspace{-5mm}

\maketitle

\begin{abstract}

Finding hot topics in scholarly fields can help researchers to keep up with the latest concepts, trends, and inventions in their field of interest. Due to the rarity of complete large-scale scholarly data, earlier studies target this problem based on manual topic extraction from a limited number of domains, with their focus solely on a single feature such as coauthorship, citation relations, and etc. Given the compromised effectiveness of such predictions, in this paper we use a real scholarly dataset from \emph{Microsoft Academic Graph} \cite{sinha2015overview}, which provides more than 12000 topics in the field of Computer Science (CS), including 1200 venues, 14.4 million authors, 30 million papers and their citation relations over the period of 1950 till now. Aiming to find the topics that will trend in CS area, we innovatively formalize a hot topic prediction problem where, with joint consideration of both inter- and intra-topical influence, 17 different scientific features are extracted for comprehensive description of topic status.

By leveraging all those 17 features, we observe good accuracy of topic scale forecasting after 5 and 10 years with $R^2$ values of 0.9893 and 0.9646, respectively. Interestingly, our prediction suggests that the maximum value matters in finding hot topics in scholarly fields, primarily from three aspects: (1) the maximum value of each factor, such as authors' maximum h-index and largest citation number, provides three times the amount of information than the average value in prediction; (2) the mutual influence between the most correlated topics serve as the most telling factor in long-term topic trend prediction, interpreting that those currently exhibiting the maximum growth rates will drive the correlated topics to be hot in the future; (3) we predict in the next 5 years the top 100 fastest growing (maximum growth rate) topics that will potentially get the major attention in CS area. All our findings are further demonstrated through an online visualization system.

\end{abstract}


\IEEEpeerreviewmaketitle

\section{Introduction}

Scientific topics are highly dynamic with breakthrough technologies promoting established domains and meanwhile stimulated the emergence of new research territories. This is especially the case in the field of Computer Science, where, as reported statistically, new technologies are constantly born at an extremely rapid growth rate of around 22\%. Consequently, it makes it impossible for researchers from that specific field to keep up with all the latest concepts, trends, and inventions in a reasonable amount of time. Very often, they may easily get lost in those massive emerging scholarly topics when attempting to seek for the topics that can potentially draw the major attention in the area of Computer Science in the next several years. Similar phenomenon also holds in a large number of other scholarly fields such as physics, biology, chemistry and etc. Under such circumstance, it is desirable to have a mechanism that can effectively help researchers find, in those fields of their interests, the hot topics in the future.

Literally, finding hot topics is closely related to topic tendency prediction, topic formation analysis as well as topic development forecasting. While those lines of work have been intensively studied in traditional social networks\cite{jang2013discovering,Lin2011The,saha2012learning,Zhu2014Tracking,guo2015social,chen2012towards}, there has been little attention given to the seek of hot topics in scholarly fields. Among those very few that indeed have tie to finding hot research topics, existing works target this problem by borrowing from traditional social network the approaches that adopt manual topic extraction from a limited number of domains, with their focus solely on coauthorship, citation relations or other types of a single aspect. For example, Hurtado $et$ $al.$ \cite{Hurtado2016Topic} extract topics from a collection of documents from both the domains of Data Mining and Machine Learning and forecast topic trends in the near future. Adopting the idea of community partitioning in social network, Qian $et$ $al.$\cite{qian2014topic} aims to reveal the underlying process and reasons from topic formation and development of scholarly networks.

Despite the delightful predictability of such predictions in conventional social networks, it encounters compromised effectiveness when applied directly to scholarly networks due to the intrinsic difference between social and scholarly networks, primarily for three reasons. First, in social networks, topics are always extracted from texts by calculating standard textual similarity, which, unfortunately, leads to poor accuracy performance when adopted in scholarly networks. This is because in an academic paper, commonly used words with ambiguous meanings are more likely to be selected due to their frequent appearance while the technical terms with significantly lower occurrence frequency can be easily missed out. Second, the hierarchical structure of social topics, which means some small topics belong to the same high level theme, is traditionally obtained by clustering the small topics. However, very often academic topics extracted from a limited number of domains fail to provide sufficient quantity of topics to do the clustering, and for the first aspect, the extracted topics cannot provide precise information to divide the hierarchy. Third, in contrast with social topics that can be well extracted from contents, scholarly topics turn out to be more complicated, containing various factors (e.e., paper, author, citation relationship, and etc.), and the joint influence of which on topics, however, has been ignored to a large in current related literature. The underlying solution to the three problems requires a large-scale scholarly dataset that can provide complete scholarly information, especially the topic information of papers, but unfortunately, these datasets are absent in the previous work. In light of those difficulties and limitations, in the present work we use a real scholarly dataset from \emph{Microsoft Academic Graph}\cite{sinha2015overview} with more than 12000 topics, 14.4 million authors, 30 million papers, and 1200 venues in the CS field, which also contains the hierarchy structure of topics and citation relationship between papers. By investigating all the topics of CS field, our goal is to find out the potential hot topics in the future.

\textbf{Contributions} In this work, we formalize the problem of scientific factors extraction which can depict the present status and influence the future development of a topic, and finding hot topics in the future as a topic trend prediction problem. After measuring all the 12464 practical topics in the dataset which comprehensively covers all areas of the CS field, we visualize the hierarchical structure of academic topics, and jointly consider two types of factors that describe the present status and growth potential of topics: \emph{external factors}, which depict the mutual influences between different topics, and \emph{internal factors}, which measure influences imposed by the intra-topic components, such as papers, authors and related venues, etc. Taken altogether both the external and internal factors, we extensively extracts 17 different scientific features to evaluate the status of a topic and perform topic trend prediction.

By leveraging all the 17 aforementioned features, we are able to predict the topic scale after 5 years with an $R^2$ value of 0.9893, which stays at 0.9646 even after 10 years, meaning that we can accurately predict the future trend of scholarly topics. We precisely investigate what role a variety of factors play in the prediction, of which the performance can be improved by adopting the idea of a multilayer neural network. Interestingly, our prediction results disclose that \textbf{the maximum value matters in finding hot topics in scholarly fields}, which can be further unfolded into the following three aspects:

(1) The maximum values of each factor, such as authors' maximum h-index and largest citation number can provide three times the amount of information than their average values, which reveal counterparts. Alternatively, this infers that whether the topic will be hot in the future closely relies on the highest level elements rather than average research level in this topic.

(2) The external factors which, as neglected by a large body of current literature, depict the interrelation of topics, and exhibit significant predictability in the long-term prediction. It means that topics with maximum growth rates now serves as the most crucial factor to drive the correlated topics to be hot in the future, and this effect becomes more apparent as the time span gets longer.

(3) By our predictions, we find out, in next 5 years, the top 100 fastest growing topics (with the maximum growth rate) that can potentially draw the major attention in the CS field.

To facilitate topic prediction for users, we develop an online visualization system named TopicMap, where both the statistical and predictive information of each topic of interest are displayed to users. Overall, our findings unveil important factors for scholarly topic trend prediction and forecast the top 100 hottest topics in the future, and, ultimately, can effectively help researchers keep up with the frontiers of science and technology.

The remainder of the paper is structured as follows. Section 2 introduces our problem formulation provides some key definitions. Section 3 presents analysis of factors we extract from topics. Future trend predictions and factor importance analysis are reported in Section 4. Section 5 reviews existing work in topic tendency forecast, and Section 6 gives concluding remarks.

\section{Problem Definition}

Traditionally, the task of topic tendency prediction based on the topics extracted from texts, or apply evolving models to a specific set of papers. One fatal weakness of this approach is that the data quantity is very small, and the factors used are too monotonic. Limited by the integrity of datasets, past works can only analyze a few of factors that can influence the topic trend, or even one factor's time series prediction. However, the topic development is subject to the combined effects of various factors in the topic, and topics will influence each other at the same time. In order to better predict the trend of the topic, we formalize two main problems, namely scientific factor extraction and topic trend prediction.

\textbf{Problem 1: Scientific Factor Extraction}. The goal is to examine factors that can influence the future development or the factor that can show the present status of a topic. Including the co-evolution relation between the different factors and how these factors influence the future trend.

\textbf{Definition 1: External Factor}. The External Factor is the factor which describes the mutual influences between different topics, including \emph{k-core} analysis of topic network and driving effect between similar topics.

\textbf{Definition 2: Internal Factor}. The Internal Factor is the factor which describes the information of intrinsic elements of topics, such as papers and authors related to this topic.

\textbf{Problem 2: Topic Trend Prediction}. The goal is to regard the scale prediction as a regression problem. Given the factors' value of topic $T$ at time $t$, the problem is to predict the paper number $N$, which means the size of this topic, at the time $t+\Delta t$ .

The major novelty of this approach lies in two ways. First, in contrast to the small set of topics, we examine 12464 topics of the CS field, containing more than 14.4 million authors and 30 million papers. Furthermore, compared to the general approaches to extract topics from paper texts, which always generate many ambiguous topics, topics we used are all practical topics after precisely classifying so that more convincing than the traditional extracted one.

Second, fundamentally different from the topic trend prediction with time series of a single factor, we jointly consider external and internal factors affecting the trend of topics, and make accurate predictions about the future trend of topics. Whereas many topic tendency predictions typically employ specific evolving models, the chief advantage of our formulation is its general applicability to a variety of real-world tasks, such as popularity prediction \cite{shen2014modeling}, expert findings \cite{zhang2007expert} and prediction of social network\cite{asur2010predicting}.

\section{Scientific Factor Extraction}

\begin{figure*}[!htbp]
   \begin{minipage}[b]{.49\linewidth}
     \centering
     \includegraphics[width=3.5in]{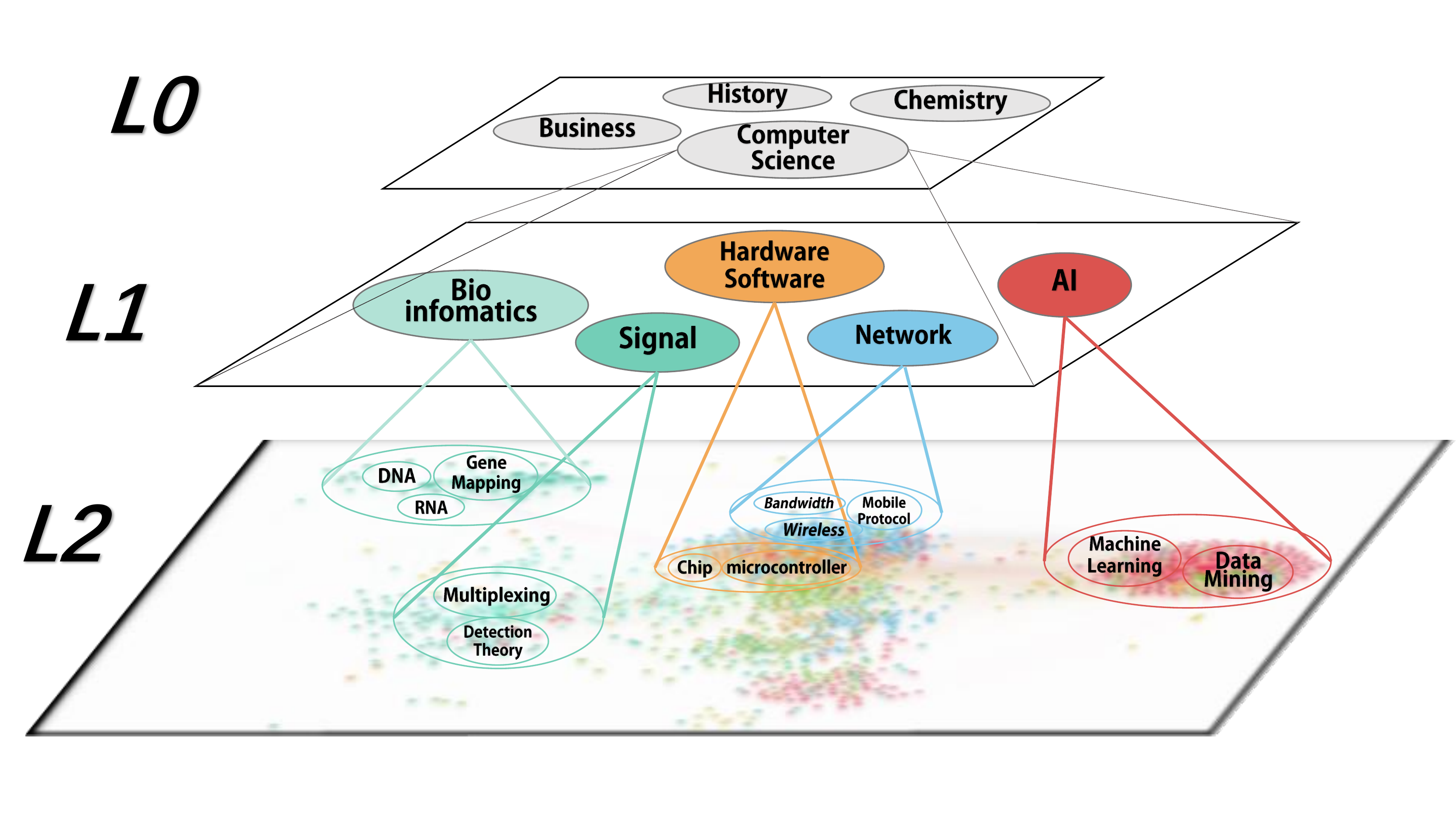}
     \text{(a) Topic Layer}
     \label{fig-layer}
   \end{minipage}
   \hfill
   \begin{minipage}[b]{.49\linewidth}
     \centering
     \includegraphics[width=3.5in]{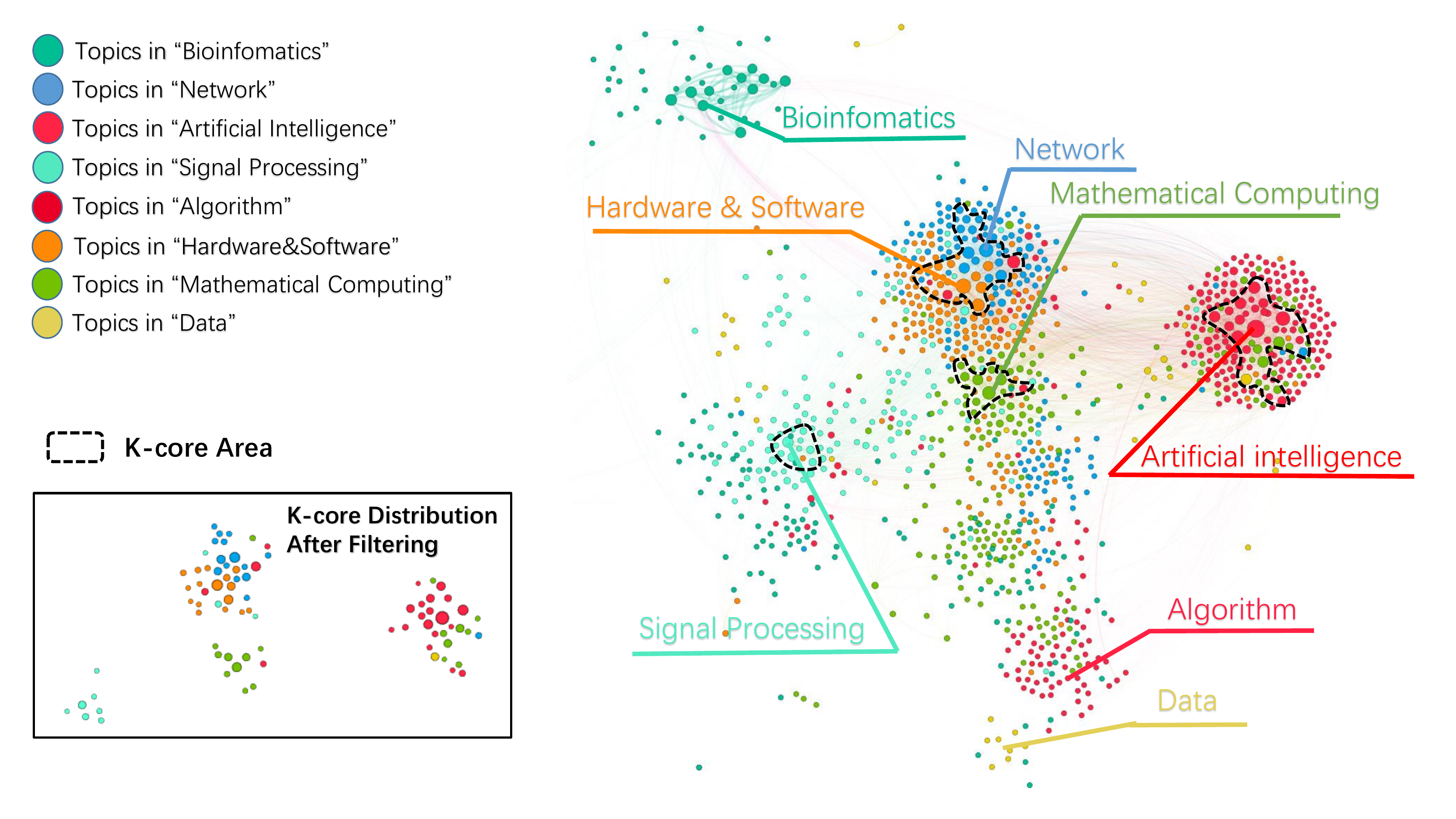}
     \text{(b) Topic Interrelation}
     \label{fig-topicmap}
   \end{minipage}
   \vspace{-2mm}
   \caption{(a)\textbf{Topic Layer.} $L0$ topics represent the basic domains of academia. $L1$ topics are basic categories of $L0$ topics. $L2$ level contains more specialized areas of $L1$ topics and the small circles in it are $L3$ topics. (b)\textbf{Topic Interrelation.} The top 10\% large topics of all $L3$ topics.Each circle means a topic and its color shows the category it belongs to. The Radius of circles represent topic The lines between topics means a connection of two linked topics. The small graph shows the cores' structure filtered by the big Topic Map.}
   \label{fig-structure}
   \vspace{-6mm}
\end{figure*}

Scholarly topics are jointly influenced by many factors. To predict the future trend of a topic, it is natural to find these factors that drive a topic to trend. In other words, we aim to find the co-evolution behavior of the factors and topic trends. Overall, these factors can be categorised as external factors, which depict the mutual influences between different topics, and internal factors, which measure influences imposed by the intra-topic components, such as papers, authors and related venues, etc. In light of this idea, we propose some inter- and intra-toptical features that can describe the growth potential of a topic, and depict their relationship with the topic future trend. Table I lists all the factors, as well as the correlation coefficients between the factors and the topic scale, which equals to the paper number in the topic after time interval of $\Delta t=1$ year, $\Delta t=5$ years and $\Delta t=10$ years, respectively.


\subsection{External Factors that Influence Topic Trend}

We first examine the mutual influence between topics. We call these kind of influences as \textbf{External Factors} that influence trend of topics.

\subsubsection{Topic Hierarchy Structure}

For the topic information is hard to get, existing work often uses standard textual similarity, such as latent Dirichlet allocation (LDA)\cite{blei2003latent}, to classify academic topics, but the ambiguous extracted topics cannot provide precise information to divide the hierarchy. Unlike common datasets, the \emph{Microsoft Academic Graph} \cite{sinha2015overview} we used, which provides topic information of each paper and the topic hierarchy structure. The hierarchy of topics is shown in Fig \ref{fig-structure}(a). It contains 4 levels in the dataset, $L0$, $L1$,$L2$ and $L3$. $L0$ level represents the basic domain of the whole academia, such as \emph{Computer Science, Mathematics, Biology etc.} We choose Computer Science as our research object. The $L0$ topic contains $L1$ topics. To the Computer Science, $L1$ topics are some basic fields of the CS area, such as \emph{Network etc.} The $L2$ and $L3$ topics are not totally parents and children relationship, for some $L3$ topics are directly belong to the $L1$ topic, but most of the $L2$ topics are bigger than the $L3$ topics. $L2$ topics contain some big concepts such as \emph{Data Mining, Machine Learning etc.} and $L3$ topics are more specific domains such as \emph{5G, Topic Model etc.} In Fig. \ref{fig-structure}(b), we draw the top 10\% large topics (according to its paper number) of $L3$ topics. Each circle represents a topic and the radius of the circle proportionate to the paper number in this topic. If two topics contain same papers, an edge will form between two topics, and the weight of this edge means the total number of papers shared by two topics. Since there are too many edges, we set a weight threshold and filter the edges whose weight is less than 500 to make the graph more clear. Then remaining edges form a strong correlation among the topics. We cluster the topics according to their similarity, paint different themes with different colors, and name clusters with top size topics in them. 

\subsubsection{k-core Analysis}
In previous work, \emph{k-core} analysis is mainly used in coauthorship network. In our research, to know the most basic structure of CS field, we first to adopt the idea of \emph{k-core} to extract the skeleton of the topic network. A \emph{k-core} is the maximum subgraph where all vertices have a degree of at least $k$. In each cycle, the vertices whose degree is less than the threshold $k$ will be deleted from the graph, and edges connected to this vertex will be deleted at the same time. The number of remaining topics after \emph{k-core} processing changes with the value of threshold $k$. We measure the \emph{k-core} feature of each topic, which equals to the value of $k$ when this topic is filtered from the network. Larger \emph{k-core} means that this topic has more connections with other topics and in a more central position in the topic network. From Fig. \ref{fig-structure}(b), the max value of $k$ we can set is 59, which means if $k$ is larger than 59, all vertices will be filtered. We can see that some core topics are at the centers of each cluster. The subgraph after \emph{k-core} filtering is as shown in Fig. \ref{fig-structure}(b). We can see that these topics are the foundations to support the entire CS field.

\begin{figure}[!t]
   \begin{minipage}[b]{.49\linewidth}
     \centering
     \includegraphics[height=1.5in, width=1.6in]{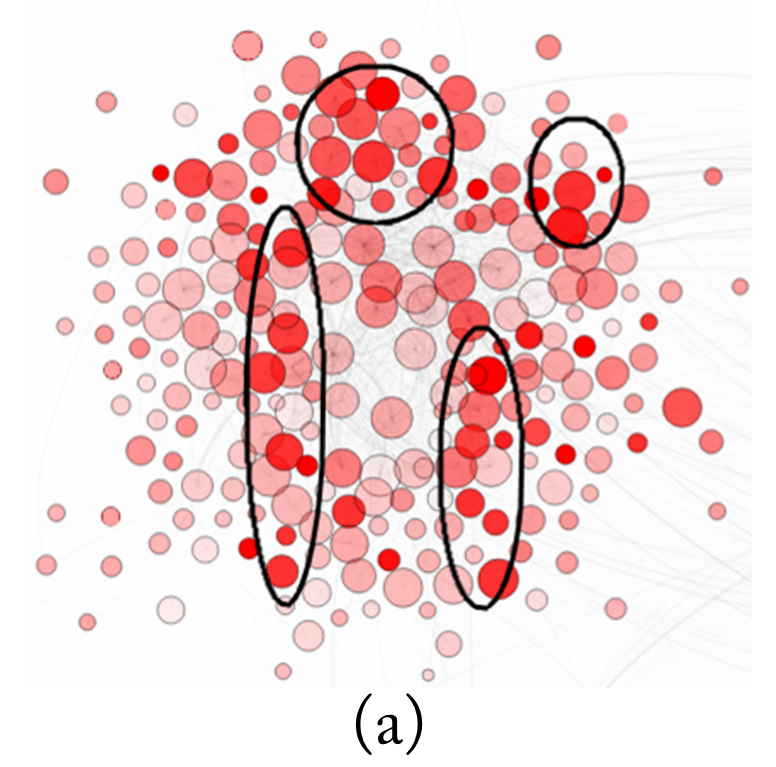}
     \label{fig-ratio-1}
   \end{minipage}
   \hfill
   \begin{minipage}[b]{.49\linewidth}
     \centering
     \includegraphics[height=1.4in, width=1.6in]{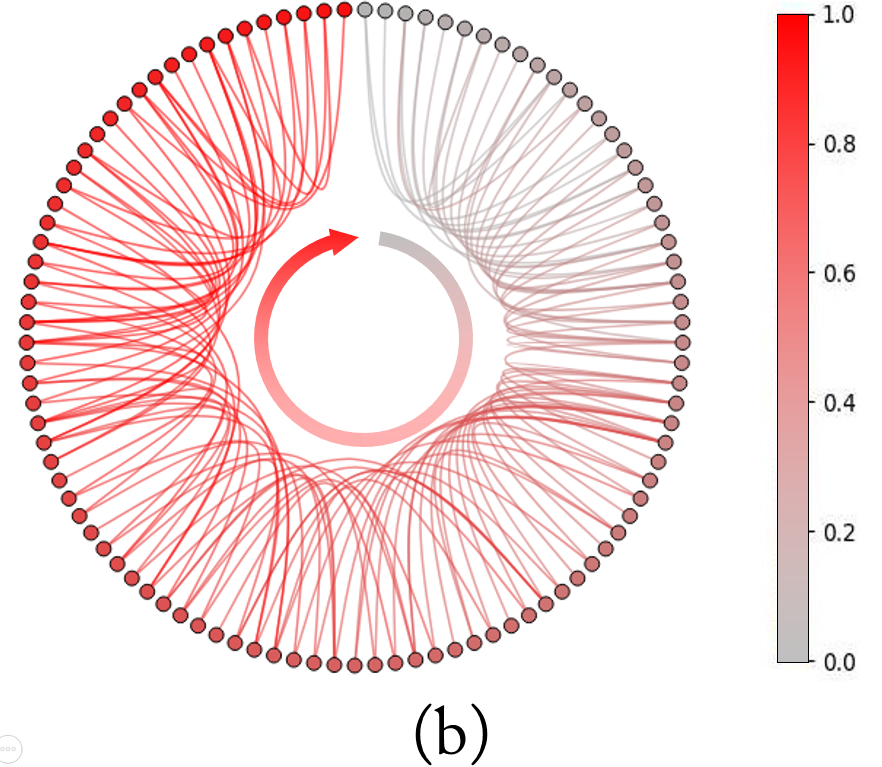}
     \label{fig-ratio-2}
   \end{minipage}
   \vspace{-5mm}
   \caption{\textbf{Topic Heat Relationship.} (a)Hot Areas (b)Growth Rate Relationship. Red means high growth rates and gray means low growth rate. Links between circles represent the most possible connections between different growth rate sets.}
   \label{fig-ratio}
   \vspace{-5mm}
\end{figure}
\subsubsection{Mutual Influences Quantification}

For the topics in the whole field are not isolated, the connection between topics will influence the their future trend. We represent the recolored version in Fig. \ref{fig-ratio}(a), where colors of topics having higher growth rates are denser. We can see the hot topics form some hot areas, which means topics having similar growth rates are more likely to have closer relationships. We call one topic's related topics as friend topics, and most of the friend topic's growth rates are similar to the present one. Fig. \ref{fig-ratio}(b) illustrates this relationship directly. After normalizing the growth rates of topics, we split growth rates into 100 slices and each circle in the graph represent a set of topics with similar growth rates. Red circles mean high growth rate sets and gray circles means low growth rate sets. Links between different sets represent the topics in these two sets have a strong connection. The graph shows that strong connections always happen between similar growth rates and do not have big spans on the graph, which reveals that the one topic's growth rate is similar to the growth rate of its friend topics.

We quantify this effect which help us predict the future trend of topics. For we have got the weight of edges between any two topics which equals to the number of common papers these two topics containing. After screening out the top 5 topics which have the closest relationships with the present topic, we regard these topics as neighboring topics of the present one. Then we define the \emph{interaction-growthnum-ave}, which is the average of \emph{increase-num} of neighboring topics. The \emph{interaction-growthnum-ave} can represent the community's growth rate and show the growth potential of this small community. Furthermore, if one topic in the neighboring topics suddenly get much attention, this effect may radiate to the present topic and make it get attention, too. So we calculate the max \emph{increase-num} of the neighboring topics as \emph{interaction-growthnum-max} and this element can quantify the driving effect of a neighboring hot topic.

\vspace{-3mm}
\subsection{Internal Factors that Influence Topic Trend}
\vspace{-2mm}

Upon analysis of external factors, now we turn to the describing different elements such as papers, authors or venues within a topic. And we call these factors the \textbf{Internal Factors}. Furthermore, after analyzing each factor, we mining the potential information of it to increase its interpretability to the topic future trend.

\subsubsection{Paper Factor}
To predict the future trend of a topic, paper is the essential factor. The famous papers make the base of its topic and attract more attention from researchers to focus on this topic, and the new papers attracted by the old paper give this topic more impact. This forms a circle to make the large and important topics have higher growth rates than the smaller topics. The number of paper in this topic is the fundamental elements of the topic. We define \emph{paper-num} to represent the number of papers in this topic at a certain year. Papers' citation information can also represent a topic's impact. One highly cited paper may lay a foundation to the related topic and open a new era of technology wave, so we define \emph{citation-max} of a topic as the maximum citation number of papers in this topic. For the similar reason, if papers' average citation in one topic is higher than another topic, this topic will obviously obtain more attention, which will increase the size of this topic in the future, so we define \emph{citation-ave} to indicate the average citations of all the papers in this topic.

\subsubsection{Author Factor}
The relationship between the number of authors and the development of academic topics cannot be ignored. Owing to the increasing participation of authors who focus on one topic, this topic could be more intriguing and be more influential in the near future. So finding a reliable method to determine who have the greater influence on the development of the topic is very significant and meaningful. In this paper, we utilize two kinds of metrics to simplify this problem.

\textbf{PART I: h-index, a standard author-level metric}

h-index was suggested in 2005 by Hirsch \cite{hirsch2005index} for attempting to measure both the productivity and citation impact of the publication of scientists or scholars. In this paper, the metric is also one of the very important factors of scientists in influencing the development of academic topics. The definition of h-index is that a scholar with an index of $h$ has published more than $h$ papers each of which at least has $h$ citations up to now. Therefore, the index can relatively reflect the influence of a scholar according to his citations and papers. We define \emph{author-hindex-ave} and \emph{author-hindex-max} to indicate the average and maximum h-index of all the authors in the topic, to show the average and maximum research level of researchers in this field.

The next factor we extract from the author part is \emph{author-hindex-var}, which means the variance of the h-index. Just like the researches in the socialism, the gap between the wealthy and poor will affect a society's development, so that the differences between high h-index authors and lower ones also show some influence on the growth of the whole topic. The \emph{author-hindex-var} show the influence in mathematical form.

\textbf{PART II: PCType, a self-designed author-level metric}

However, the index cannot distinguish the author who ever published a paper with great numbers of citations but after then the papers published were all footy with few citations. For example, two scholars both published 99 low-cited papers before, but one of them later published a paper with great influence. The result is that they both have low h-index. So the absolute number of the citation number and paper number cannot be ignored. In this part, we propose PCType, which takes the paper number and citation number into consideration, for mining more meaningful information of scholars.

Based on our assumption, if some authors with higher citation numbers and paper numbers, who is also more insightful in our opinions, take part in one topic, it means they admit the potential or the importance of this topic. Therefore, in this part, we mainly focus on the real distribution of \emph{authorCitation-num} as citation number and \emph{authorPaper-num} as paper number of each author and define four kinds of author due to the distribution as follows:

\vspace{-2mm}
\begin{table}[h]  
\center 
\begin{tabular}{|c|c|c|}
\hline               
authorTpye & \emph{authorPaper-num}  & \emph{authorCitation-num}  \\  
\hline
PHCH  & high & high \\
PHCL  & high & low \\  
PLCH  & low & high \\  
PLCL  & low & low \\  
\hline  
\end{tabular}  
\end{table}  
\vspace{-3mm}

\begin{figure}[!t]
   \begin{minipage}[b]{.49\linewidth}
     \centering
     \includegraphics[width=1\textwidth]{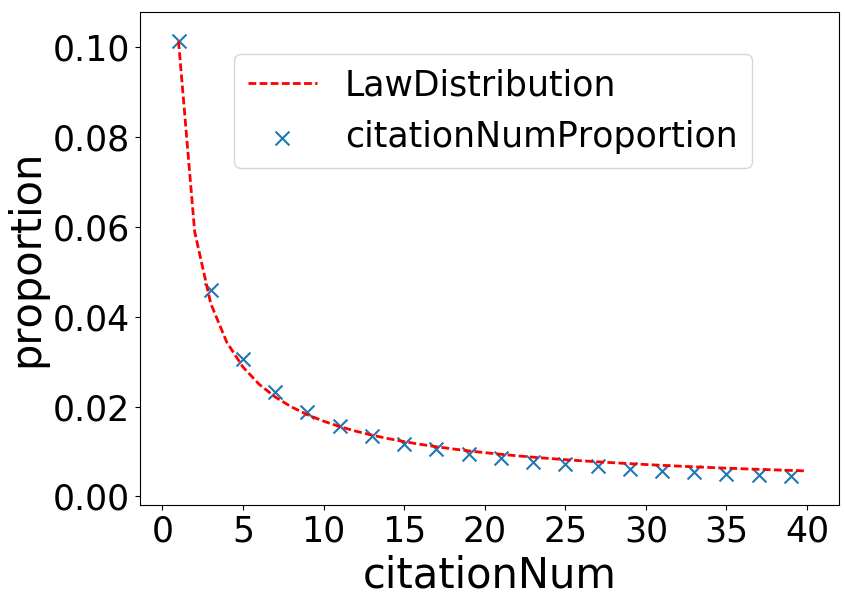}
     \text{\footnotesize{(a)}}
     \label{subfig4-1}
   \end{minipage}
   \hfill
   \begin{minipage}[b]{.49\linewidth}
     \centering
     \includegraphics[width=1\textwidth]{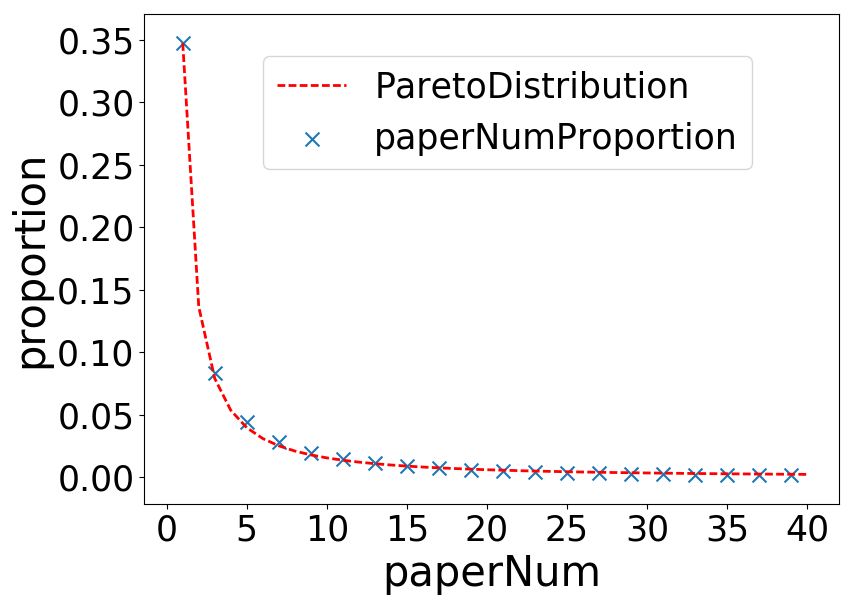}
     \text{\footnotesize{(b)}}
     \label{subfig4-2}
   \end{minipage}
   \vspace{-7mm}
   \caption{\textbf{Real Distribution.} of (a)authorCitation-num proportion and (b)paper-num proportion, based on the total information of authors and papers from 1900 to 2016 in whole CS field}
   \label{fig4}
   \vspace{-5mm}
\end{figure}

In this part, we utilize the dataset about information of all papers and authors in the CS field provided by \emph{Microsoft Academic Graph}\cite{sinha2015overview}. In order to obtain the break point for distinguishing the high level and low level of authorCitation-num and authorPaper-num , we take the following three steps:

\textbf{Step 1: Real distribution.}
As Fig.~\ref{fig4} shows, we draw two graphs, including the real proportion of authorCitation-num with citation number increasing and standard pareto distribution, and the real proportion of authorPaper-num with paper number increasing and modified pareto distribution based on the total citation-num and author-num from 1900 to 2016.

\textbf{Step 2: Probability density function.}
The standard pareto distribution is to describe the distribution of a random variable, the probability that X is greater than some number x is given by
\vspace{-1mm}
$$\overline{F}(x) = Pr(X>x) =
\begin{cases}
(\frac {x_m}{x})^\alpha& {x \geq x_m}\\
1& {x < x_m}
\end{cases}$$
\vspace{-1mm}
where x$_m$ is the minimum possible value of X, and $\alpha$ is a positive parameter called pareto index. Hence the probability density function of X followed is
$$f_X(x) =
\begin{cases}
\frac {\alpha x_m^\alpha}{x^{\alpha+1}}& {x \geq x_m}\\
0& {x < x_m}
\end{cases}$$
\vspace{-1mm}
As Fig.~\ref{fig4} presents, we get the conclusion that probability density function of paper-num obeys the pareto distribution with pareto index $\alpha_{paper}$ = 0.347. And for the probability density function of citation-num, it obeys the law distribution with attenuation coefficient $\beta_{citation} = 0.782$
\vspace{-1mm}

$$f_X(x) =
\begin{cases}
(\frac {x_m}{x})^\beta& {x \geq x_m}\\
0& {x < x_m}
\end{cases}$$
\vspace{-1mm}

where x$_m$ is the minimum possible value of X, and $\beta$ is a positive parameter called attenuation coefficient.

\begin{figure}[!t]
\centering
\includegraphics[width=3in]{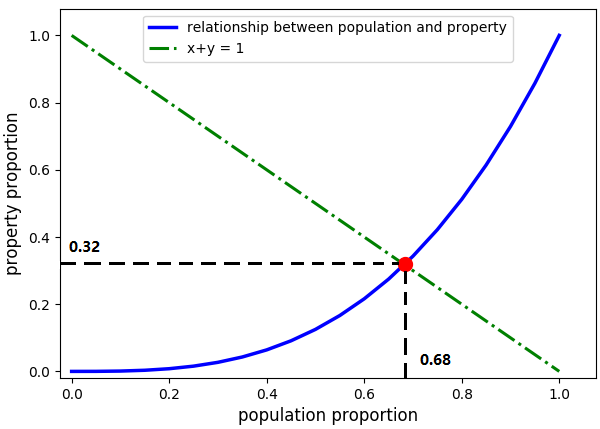}
\vspace{-4mm}
\caption{\textbf{Relationship between the proportion of population and property.} Pareto principle applied in relationship between the proportion of population and property, as 68\% of population only holds 32\% of wealth.}
\label{fig5}
\vspace{-3mm}
\end{figure}

\textbf{Step 3: Break point.}
Owing to the distribution of citation-num and paper-num obeying standard pareto distribution and law distribution, the pareto principle can be utilized to calculate the break point. The pareto principle is to describe a phenomenon that for many events, most of the effects come from little of the causes. For example, as Fig.~\ref{fig5} presents, let's define function f(x) as the proportion of total property and x as the proportion of total population form the poor to the rich, and the break point $x_{point}$ satisfies that
$$ x_{point}+f(x_{point}) = 1 $$

\begin{figure}[!t]
    \begin{minipage}[b]{.49\linewidth}
        \centering
        \includegraphics[height=1.2in,width=1\textwidth]{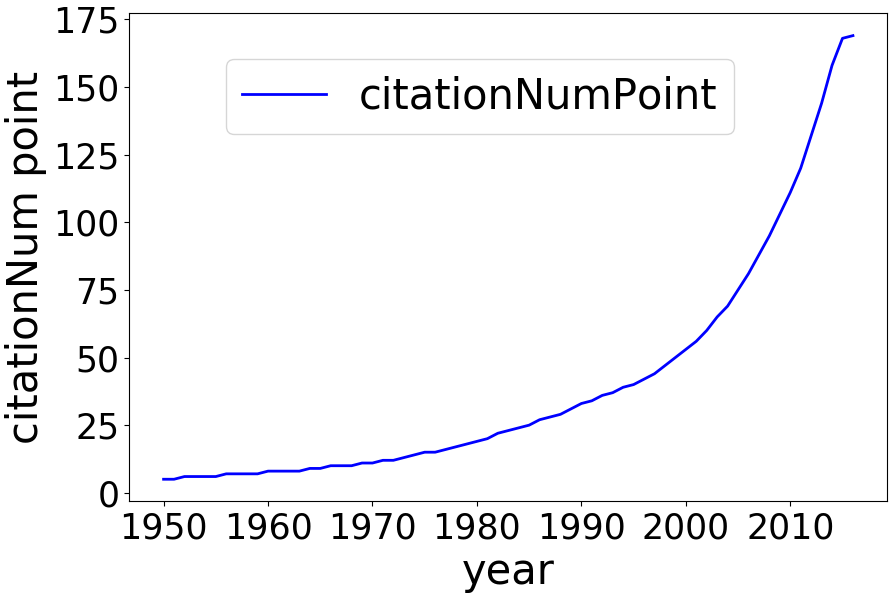}
        \text{\footnotesize{(a)}}
        \label{subfig6-1}
    \end{minipage}
    \hfill
    \begin{minipage}[b]{.49\linewidth}
        \centering
        \includegraphics[width=1\textwidth]{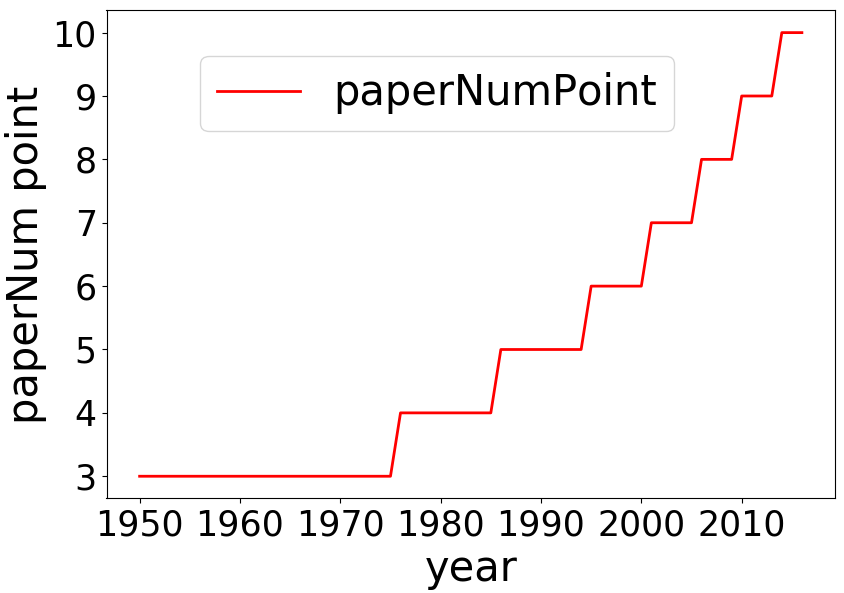}
        \text{\footnotesize{(b)}}
        \label{subfig6-2}
    \end{minipage}
    \vspace{-7mm}
    \caption{(a)The break points of citaitonNum for each year from 1950 to 2016, and (b)The break points of paperNum for each year from 1950 to 2016}
    \label{fig6}
    \vspace{-5mm}
\end{figure}

After these three steps by drawing the real distribution of both indices, deriving the probability density function, using pareto principle to get two break points. As Fig.~\ref{fig6} denotes, we calculate the two break points for each year from 1950 to 2016.

\begin{figure}[!t]
    \begin{minipage}[b]{.49\linewidth}
        \centering
        \includegraphics[width=1\textwidth]{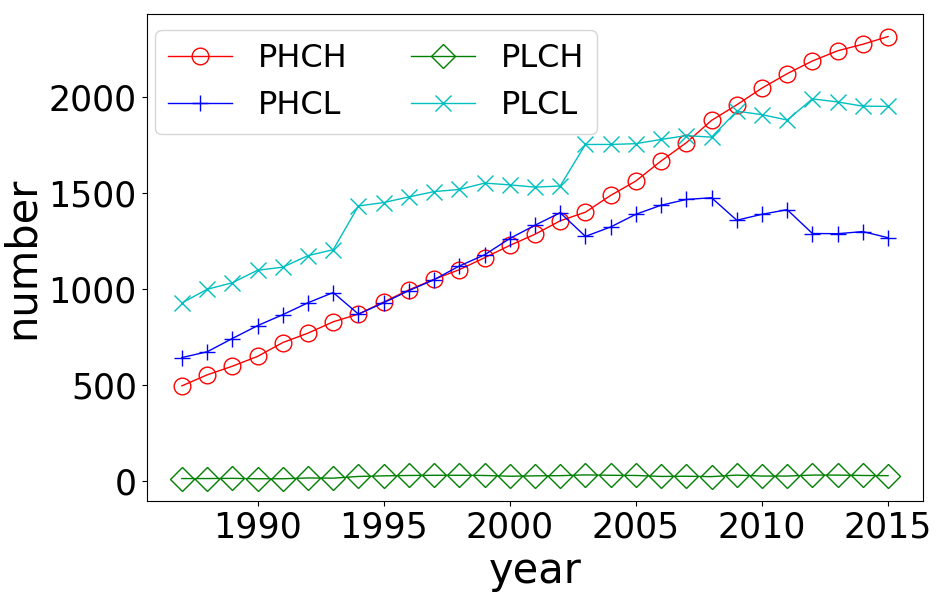}
        \text{\footnotesize{(a)}}
        \label{subfig7-1}
    \end{minipage}
    \hfill
    \begin{minipage}[b]{.49\linewidth}
        \centering
        \includegraphics[width=1\textwidth]{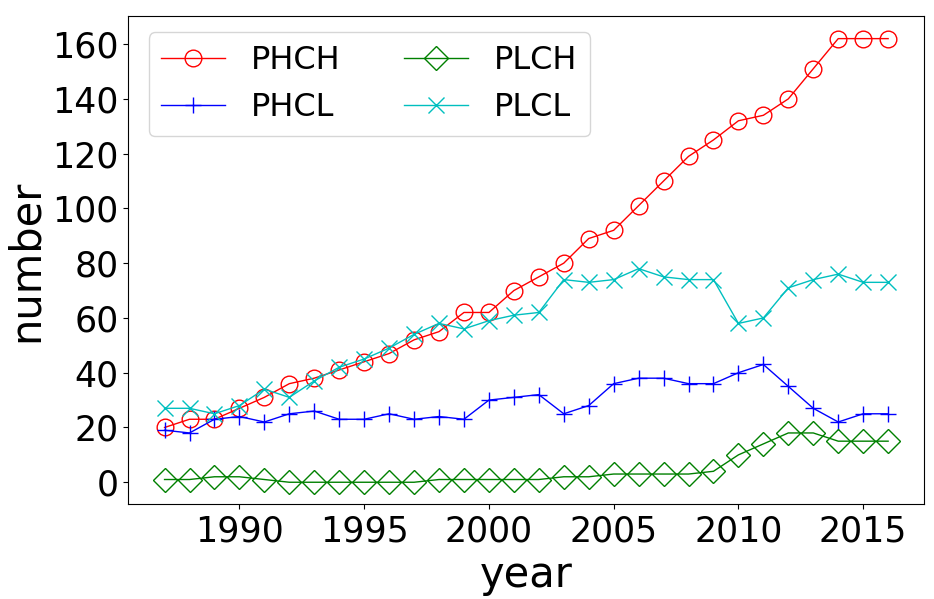}
        \text{\footnotesize{(b)}}
        \label{subfig7-2}
    \end{minipage}
    \vspace{-7mm}
    \caption{(a)The development of four types of authors about topic \emph{Coupling coefficient of resonators} and (b)The development of four types of authors about topic \emph{HAMP domain}}
    \label{fig7}
    \vspace{-6mm}
\end{figure}

Based on this result, we classify the authors into four types, PHCH, PHCL, PLCH and PLCL, and count the number for each type in each year. Our observation is that the number of PHCH is increasing steadily, the number of PHCL and PLCH is increasing with complementary trend, which means when one is increasing, the other is declining relatively. We plot two graphs about the development of four types of authors in Fig.~\ref{fig7} to illustrate the phenomenon more persuasively and clearly.

\begin{table*}[!t]
\renewcommand{\arraystretch}{1.0}
\caption{Topic Factors and Correlation Coefficients Between This Element and Topic Scale After $t$ Years}
\label{factor_table}
\centering
\begin{tabular}{c||l||l|c|c|c}
\hline
  &Feature & Definition & $cc_1$ & $cc_5$ & $cc_{10}$\\
\hline
\multicolumn{6}{l}{External Factor}\\
\hline
\multirow{3}{*}{Interaction} & \emph{k-core} & The value of $k$ in the \emph{k-core} analysis & 0.3862 & 0.3877 & 0.3944\\
& \emph{interaction-growthnum-ave} & The average value of \emph{increase-num} of neighboring topics & 0.0291 & 0.0356 & 0.0290\\
& \emph{interaction-growthnum-ave} &  The max value of \emph{increase-num} of neighboring topics & 0.0331 & 0.0381 & 0.0290\\
\hline
\multicolumn{6}{l}{Internal Factor}\\
\hline
\multirow{3}{*}{Paper} & \emph{paper-num}& The number of papers in this topic & 0.9927 & 0.9861 & 0.9570\\
& \emph{citation-ave} &  The average value of papers' citations in this topic & -0.0103 & -0.0007 & -0.0029\\
& \emph{citation-max} &  The max value of papers' citations in this topic & 0.3368 & 0.3413 & 0.3373\\
\hline
\hline
\multirow{5}{*}{Author}
& \emph{author-hindex-ave} &  The average value of authors' h-index in this topic & 0.0688 & 0.0629 & 0.0637\\
& \emph{author-hindex-max} &  The max value of authors' h-index in this topic & 0.3580 & 0.3691 & 0.3811\\
& \emph{author-hindex-var} &  The variance of authors' h-index in this topic & 0.0542 & 0.0486 & 0.0500\\
& \emph{author-phch-num} &  The number of PHCH authors in this topic & 0.8288 & 0.8154 & 0.7959\\
& \emph{author-plch-num} &  The number of PLCH authors in this topic & 0.5060 & 0.4824 & 0.4660\\
\hline
\hline
\multirow{3}{*}{Growth} & \emph{increase-num} &  The growth of paper number between current year and last year & 0.8885 & 0.9432 & 0.9438\\
& \emph{increase-num-ave} &  The average value of growth number in the past five years & 0.9487 & 0.9586 & 0.9558\\
& \emph{increase-num-max} &  The max value of growth number in the past five years & 0.9381 & 0.9385 & 0.9294\\
\hline
\hline
\multirow{3}{*}{Venue} & \emph{venue-num} &  The total number of venues in this topic & 0.7054 & 0.6767 & 0.6511\\
& \emph{venue-distinct-num} & The number of distinctive venues in this topic & 0.5669 & 0.5616 & 0.5550\\
& \emph{venue-index-ave} & The weighted average of the \emph{venueIndex} of venues appeared in this topic. & 0.0123 & 0.0280 & 0.0528\\
\hline
\end{tabular}
\vspace{-5mm}
\end{table*}

\subsubsection{Growth Factor}
The growth trends of academic topics are not as volatile as stocks, so the growth rate in the past several years may affect the future trend. We define the \emph{increase-num} to represent the growth of the paper number in a topic between the current year and last year. We also calculate the average growth of the past 5 years as \emph{increase-num-ave} to show the growth constancy of this topic. Furthermore, if a topic suddenly gets a lot of attention for the new theory comes out in this field, obviously the topic will grow very fast in the following years. For this reason, we calculate the maximum value of the growth in the past 5 years in this topic and define it as \emph{increase-num-max}.

\subsubsection{Venue Factor}
The venue information of papers serves as a very important factor in assessing the topical impact, since qualities of venues differs based on the quality level of the correspondingly published papers. For each topic, we first obtain the venues encompass this topic, which means that at least one paper in this topic have previously appeared these venues before. We define \emph{venue-num} to represent the total number of venues in this topic. Among these venues, some venues are belongs to the same series of conferences or journals (e.g., ICDM conference on each year). Thus we remove the duplicate venues belonging to the same series, and get the \emph{venue-distinct-num} as the number of distinctive venues in this topic to characterize its diversity. Obviously, the more different venues appear in this topic, the more wide-ranging this topic may be, and it may get more attention in the future.

Generally speaking, the influence of the paper is proportional to the influence of the conference or the journal. We want to distinguish the impact of papers from what venues they appear, so the first task is to determine the impact of a certain conference or journal. To measure the impact of a venue $V$, we quantify the impact as \emph{venueIndex} by
\begin{equation}
venueIndex(V) = \frac{{\sum\limits_{p \in V} {citations(p)} }}{N_V}
\end{equation}
where $N_V$ is the total number of papers in this venue and $citations(p)$ is the citation number of paper $p$. One venue's \emph{venueIndex} not only reflects the average citations of all papers in this venue, but also discloses the impact of this venue. Then we define the \emph{venue-index-ave} of topic $T$ as following:
\begin{equation}
\emph{venue-index-ave(T)} = \frac{{\sum\limits_{p \in T,V} {venueIndex(V)} }}{{{N_T}}}
\end{equation}
where $N_T$ is the total number of papers in this topic, and the \emph{venue-index-ave} of topic $T$ is the weighted average of the venueIndex of venues appeared in this topic. This factor can help us to quantify the venues' impact to a topic and help us to predict the topic's future trend.

\section{Experiment}

Based on all the 17 aforementioned features, we employ a series of models to predict the scale of topics in the future. First, we use some traditional models and some ensemble models to forecast the future trend. Then we compare the importance of each feature we proposed in the previous section and make an explanation to the research results. Finally we use the multilayer neural network to enhance our prediction.

\subsection{Experimental Setup}

We perform our experiments on \emph{Microsoft Academic Graph} (MAG)\cite{sinha2015overview} which is an official and authoritative scholarly dataset containing massive scholarly information about publications such as papers, authors, conferences, fields of study and citation relationships. Around more than 12000 topics, 14.4 million authors, 30 million papers, and 1200 venues are included in the CS field. Our primary task is to predict the topic scale, which equals to the paper number in the topic, after time interval $\Delta t$. For the topic's status are described by the factors we propose in Table I, we do the time serialization to each factor of 12464 topics from 1950 to 2015. First, we extract the paper list of each topic and split the papers into various parts by their published years. When we calculate each factor at time $t$, we use the subset of papers published earlier than $t$. Finally we get the 12464 topics' time series of each features, which containing more than 800000 time samples.

\begin{figure}[!htbp]
   \begin{minipage}[b]{.31\linewidth}
     \centering
     \includegraphics[height=1in, width=1.15in]{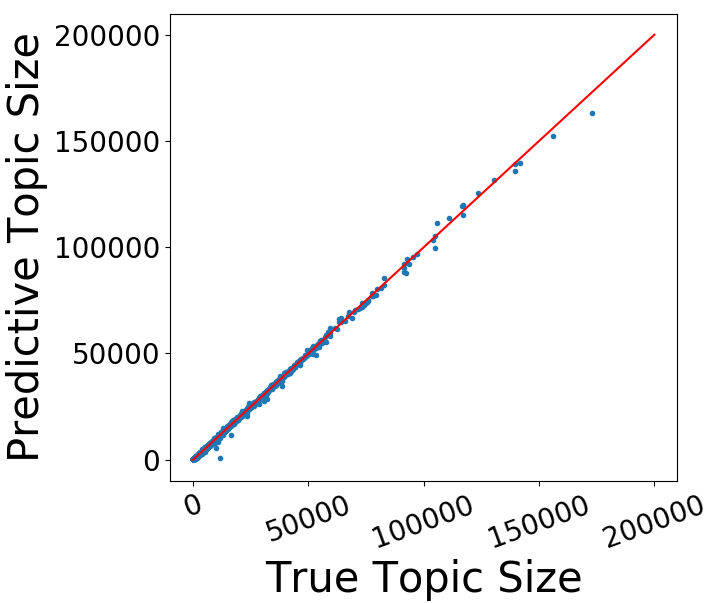}
     \text{\footnotesize{(a) $\Delta t=1$}}
     \label{subfig-1:vs}
   \end{minipage}
   \hfill
   \begin{minipage}[b]{.31\linewidth}
     \centering
     \includegraphics[height=1in, width=1.15in]{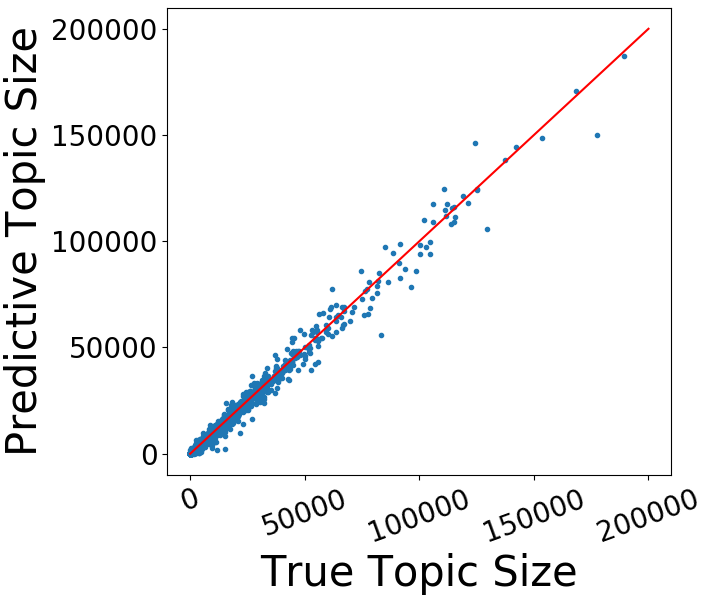}
     \text{\footnotesize{(b) $\Delta t=5$}}
     \label{subfig-2:vs}
   \end{minipage}
   \hfill
   \begin{minipage}[b]{.31\linewidth}
     \centering
     \includegraphics[height=1in, width=1.15in]{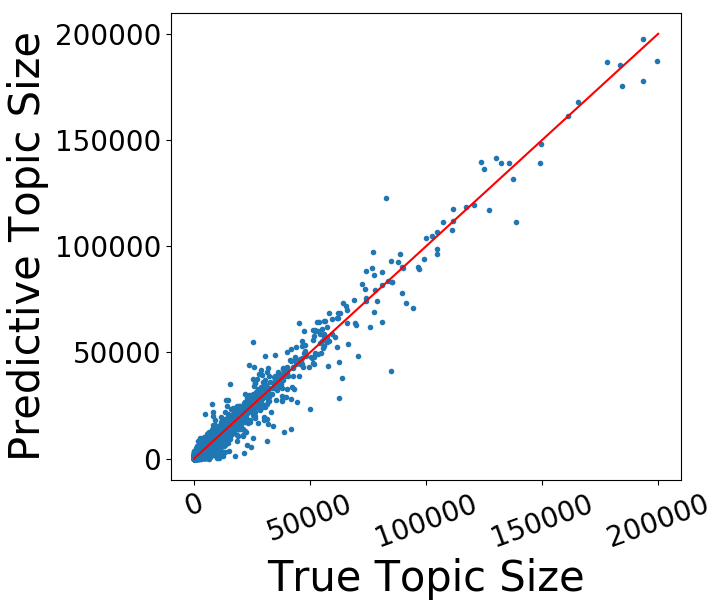}
     \text{\footnotesize{(c) $\Delta t=10$}}
     \label{subfig-3:vs}
   \end{minipage}
   \vspace{-1mm}
   \caption{\textbf{Comparison of true value and predicted value.} Prediction performance on the test dataset with $\Delta t$ = 1 or 5 or 10 years. X axis is the true values. Y axis is the predictive values. The red line denote $y=x$ which means the forecast results fit perfectly with the true values.}
   \label{fig:vs}
   \vspace{-4mm}
\end{figure}

\begin{table*}[!t]
\renewcommand{\arraystretch}{1.0}
\caption{Comparison between models predicting topic scale after $\Delta t$ Years}
\label{performance_table}
\centering
\begin{tabular}{c||c||c|c|c|c|c|c|c|c|c|c}
\hline
Criteria & Models & $\Delta t=1$ & $\Delta t=2$ & $\Delta t=3$ & $\Delta t=4$ & $\Delta t=5$ & $\Delta t=6$ & $\Delta t=7$ & $\Delta t=8$ & $\Delta t=9$ & $\Delta t=10$ \\
\hline
\hline
\multirow{6}{*}{$R^2$}
& LR  &0.9997 & 0.9983 & 0.9952 & 0.9898 & 0.9840 & 0.9749 & 0.9711 & 0.9650 & 0.9599 & 0.9487\\
& DT  &0.9984 & 0.9963 & 0.9900 & 0.9862 & 0.9748 & 0.9594 & 0.9665 & 0.9438 & 0.9375 & 0.8792\\
& RF  &0.9994 & 0.9979 & 0.9953 & 0.9921 & 0.9877 & 0.9820 & 0.9784 & 0.9701 & 0.9650 & 0.9458\\
& ExtraTrees  &0.9995 & 0.9983 & 0.9960 & 0.9926 & 0.9893 & 0.9852 & 0.9802 & 0.9732 & 0.9700 & 0.9646\\
& GBDT  &0.9995 & 0.9982 & 0.9957 & 0.9926 & 0.9880 & 0.9832 & 0.9773 & 0.9700 & 0.9634 & 0.9497\\
& BAG  &0.9994 & 0.9979 & 0.9952 & 0.9915 & 0.9882 & 0.9819 & 0.9788 & 0.9696 & 0.9663 & 0.9485\\
\hline
\hline
\multirow{6}{*}{MAE}
& LR  & 27.97 & 66.91 & 116.03 & 181.06 & 230.97 & 280.29 & 316.54 & 349.20 & 386.38 & 438.62\\
& DT  & 52.90 & 88.99 & 143.31 & 203.51 & 263.53 & 373.09 & 364.01 & 545.74 & 493.30 & 672.88\\
& RF  & 39.40 & 66.56 & 106.94 & 160.98 & 197.03 & 257.41 & 283.27 & 406.80 & 396.07 & 523.54\\
& ExtraTrees  & 37.41 & 63.89 & 102.67 & 148.97 & 183.98 & 224.09 & 265.97 & 328.66 & 349.28 & 402.29\\
& GBDT  & 50.69 & 74.41 & 111.07 & 155.07 & 197.39 & 244.56 & 282.22 & 353.16 & 374.38 & 438.84\\
& BAG & 39.17 & 66.40 & 107.03 & 160.99 & 199.60 & 256.13 & 278.73 & 397.61 & 388.92 & 501.52\\
\hline
\end{tabular}
\vspace{-5mm}
\end{table*}

\begin{figure}[!t]
   \begin{minipage}[b]{.49\linewidth}
     \centering
     \includegraphics[height=1.1in, width=1.8in]{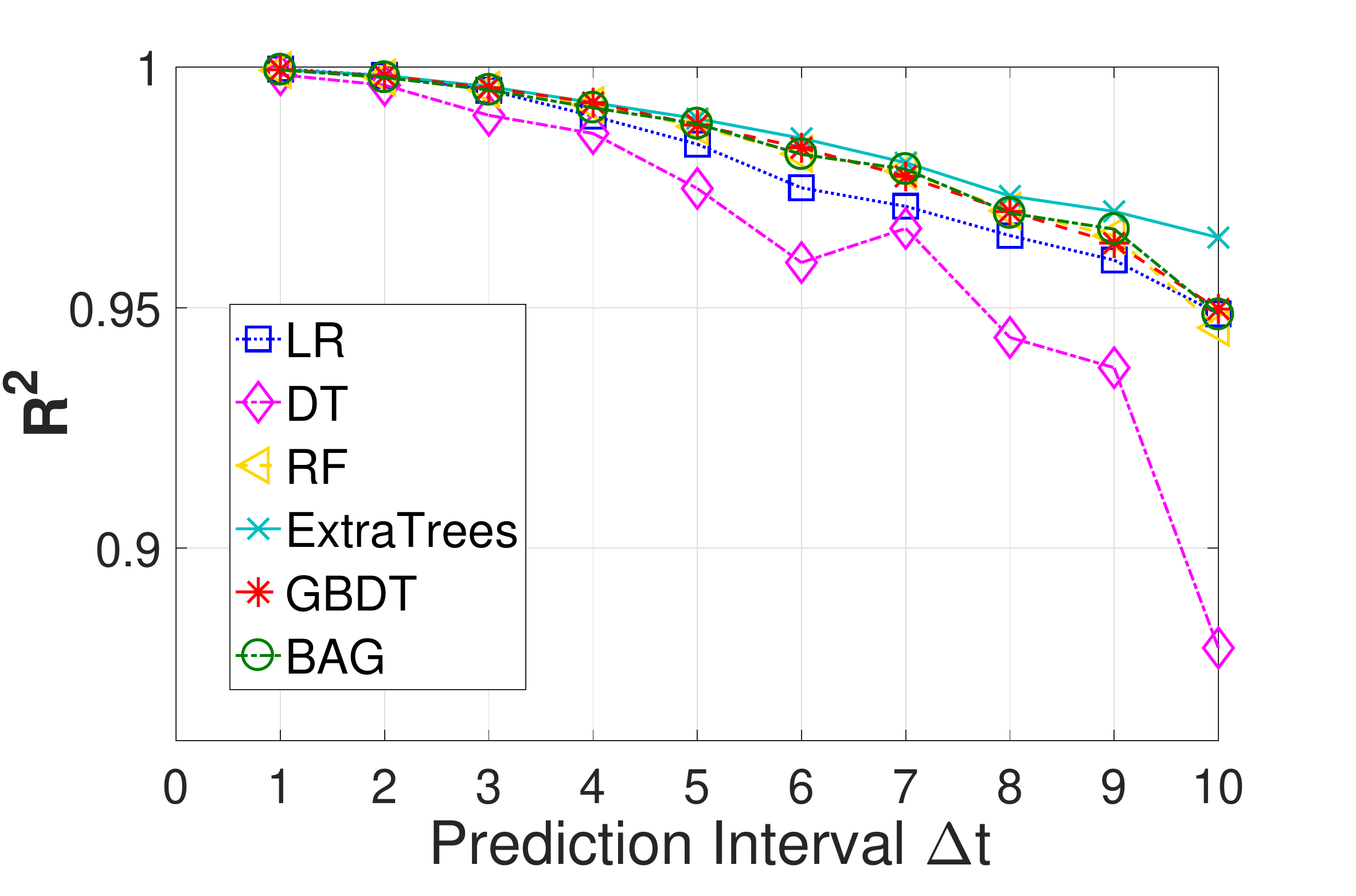}
     \text{\footnotesize{(a)}}
     \label{subfig-1:R2}
   \end{minipage}
   \hfill
   \begin{minipage}[b]{.49\linewidth}
     \centering
     \includegraphics[height=1.1in, width=1.8in]{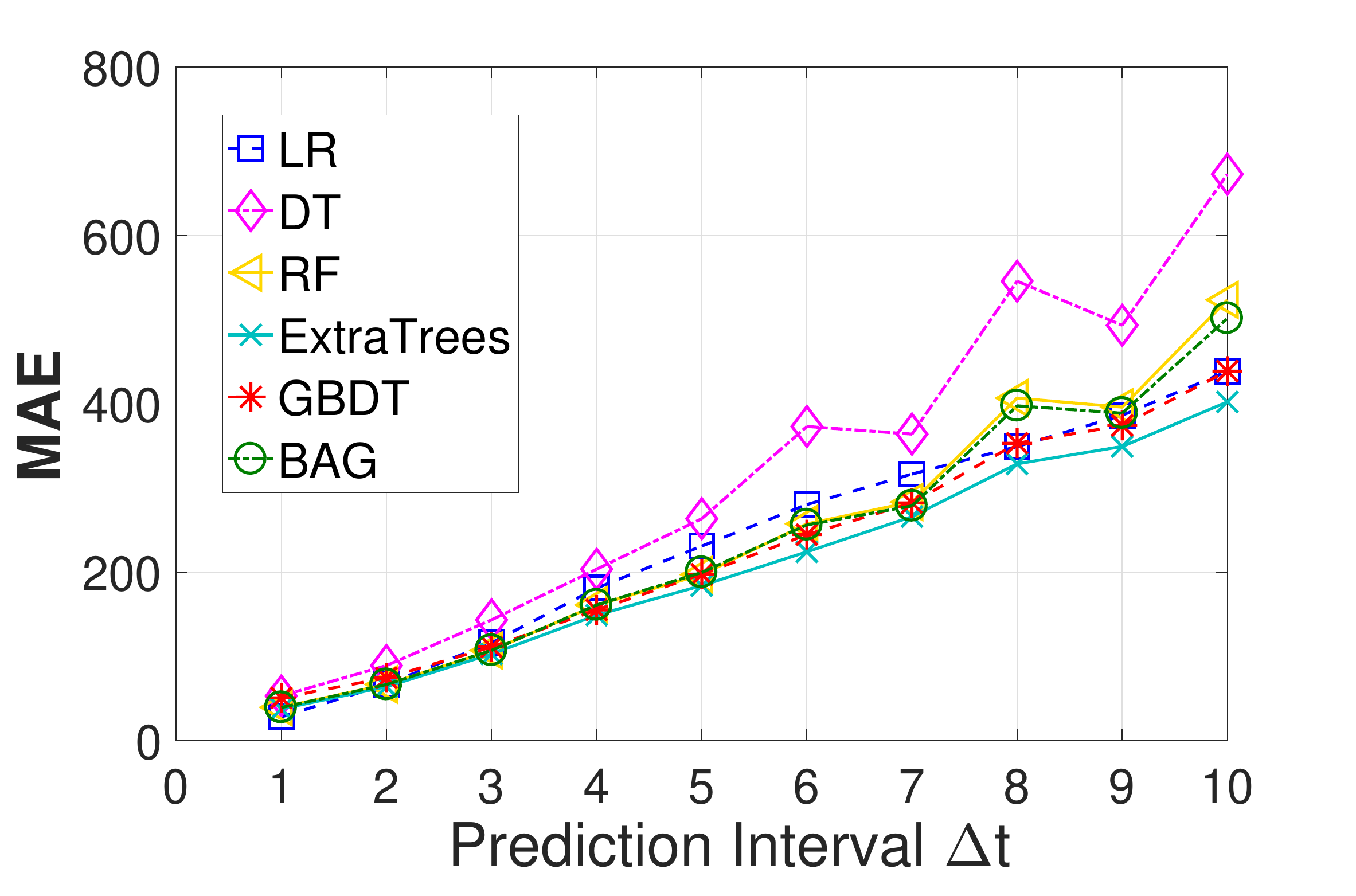}
     \text{\footnotesize{(b)}}
     \label{subfig-2:MAE}
   \end{minipage}
   \vspace{-7mm}
   \caption{\textbf{Predictive performance of two metrics.} (a) $R^2$ (b) MAE. Prediction performance of 6 models with different time interval from 1 year to 10 years.}
   \label{fig:performance}
   \vspace{-2mm}
\end{figure}

\subsection{Predicting Topics with Ensemble Model}

We use several models to predict the topic size in the future, including linear regression (LR), Decision Tree Regression (DT), Random Forest Regression (RF), Extremely Randomized Trees Regression (ExtraTrees), Gradient Boosting Regression (GBDT) and bagged decision trees (BAG). To evaluate the prediction accuracy, we compare these models by the coefficient of determination ($R^2$) and the mean absolute error (MAE). Fig. \ref{fig:performance} shows the performance of different models in terms of $R^2$ and MAE. We can see that for all the models, $R^2$ will decrease as the prediction gap becomes larger, and the MAE will increase at the same time, which both mean that our prediction has better performance in the shorter time interval. From Fig. \ref{fig:performance} we can obtain that the Extremely Randomized Trees Regression(ExtraTrees) shows the best performance among these models and achieve the $R^2$ of 0.9893 when $\Delta t=5$ and 0.9646 when $\Delta t=10$. We also get the MAE of 183.98 when $\Delta t=5$ and 402.29 when $\Delta t=10$. The detailed performance is showed in Table \ref{performance_table}.

From Fig. \ref{fig:vs} we can have an intuitive perception of the results of the forecast. We choose the ExtraTrees, which has the best performance among the models, to predict a topic's size after diverse years. The x axis represents the true values of samples in the test dataset and the y axis represents the predicted values. The red line denote $y=x$ which means the forecast results fit perfectly with the true values. Each point represents a test sample. We can see that the accuracy will be higher with the less forecast interval and our prediction performs well on topics of different scales. We can also see that for the majority of topics containing more than 10000 papers, our MAE of 183.98 when $\Delta t=5$ and 402.29 when $\Delta t=10$ shows high accuracy in predicting the future trend of scholarly topics. Through our prediction, we can determine the topic scale in the future, and find out the topics will be hot in the CS field.

\subsection{Factor Importance Comparison}

\begin{figure}[!t]
\centering
\includegraphics[width=3.5in]{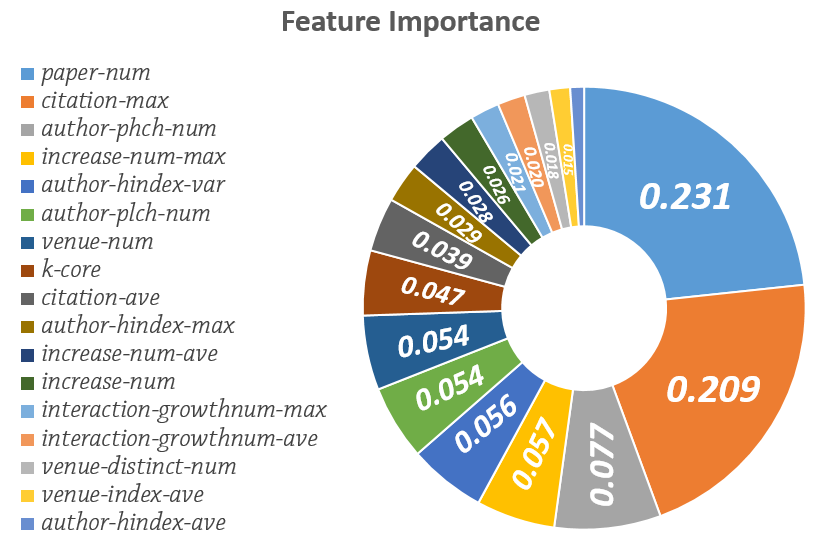}
\caption{\textbf{Feature importance analysis.} The importance of each feature measured by information gain from the extremely randomized trees regression. It illustrate that for a certain factor, the maximum value have better performance than the average ones.}
\label{fig-features}
\vspace{-5mm}
\end{figure}

For the Extremely Randomized Trees Regression has the best performance, we select this algorithm to analyze the importance of each factor we proposed. Friedman $et$ $al$ \cite{friedman2001greedy} measure the \textbf{factor importance} of factor $j$ by the average importance of it over all of the trees
\[\hat{J_{j}^2}=\frac1M \sum_{m=1}^M\hat{J_{j}^2}(T_m)\]
where $M$ represents the number of trees in the forest. The importance of factor $j$ in a single decision tree is
\[\hat{J_{j}^2}(T)=\sum\limits_{t=1}^{L-1} \hat{i_{t}^2} 1(v_{t}=j)\]
where the $L-1$ represent the number of non-terminal nodes in the $L-terminal$ node tree $T$, $v_t$ is the factor associated with node $t$, and $i_{t}^2$ is the corresponding improvement in squared-error after splitting of node $t$. Through this way we can determine the information gain by this feature and we measure the importance of each factor in our model.

We choose 5 years as the time interval and obtain the feature importance as showed in Fig. \ref{fig-features}, the factor \emph{paper-num} is most important among all the features. We also find out that the following important factor is the \emph{citation-max}, which means that the highest cited papers provide more information to the future trend of a topic. A highly cited paper may draw much attention from researchers so that the corresponding topic will be hot in the future. In contrast, the \emph{citation-ave} is not as important as \emph{citation-max}, which reveal that if a topic want to get more attention form researchers, the highest level of papers in this topic are much more important than the average research level.

Similarly, we find that the \emph{author-phch-num} and \emph{author-plch-num} are also important features. This phenomenon shows that if a topic will be hot in the future, the famous experts and appealing pioneers are indispensable. These experts inspire researches in the whole topic and lead the topic to be hot. Among all the features, we can note that the max value of each factor play more important roles than the average value. The maximum values of each factor can provide three times the information than the average values, which determine that the max values are really important for predicting the topic trend. Topics have higher max values attract more people in the academia or industry to study and apply the related technology to industry. At the same time, these topics will be hot spots in the future.

\begin{figure}[!t]
   \begin{minipage}[b]{.49\linewidth}
     \centering
     \includegraphics[height=1.0in, width=1.7in]{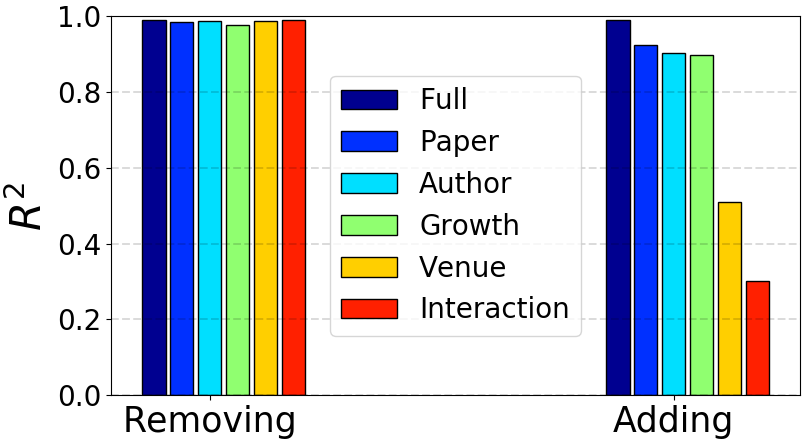}
     \text{\footnotesize{(a)$\Delta t=5$, $R^2$}}
     \label{R2-5-cmp}
   \end{minipage}
   \hfill
   \begin{minipage}[b]{.49\linewidth}
     \centering
     \includegraphics[height=1.0in, width=1.7in]{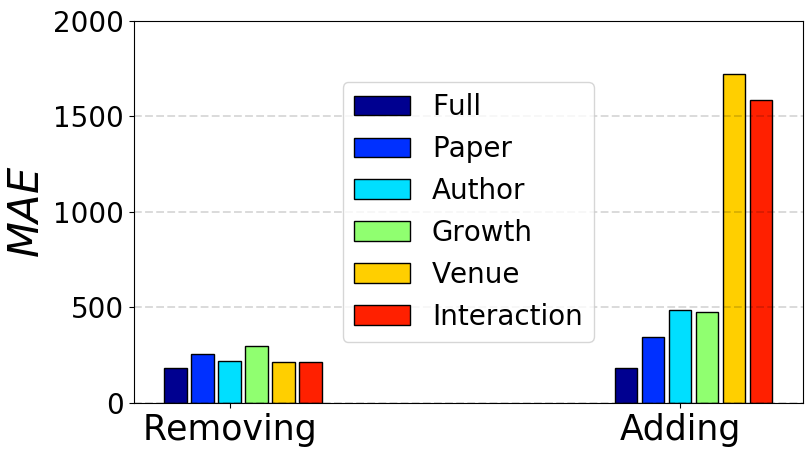}
     \text{\footnotesize{(b)$\Delta t=5$, MAE}}
     \label{MAE-5-cmp}
   \end{minipage}
   \begin{minipage}[b]{.49\linewidth}
     \centering
     \includegraphics[height=1.0in, width=1.7in]{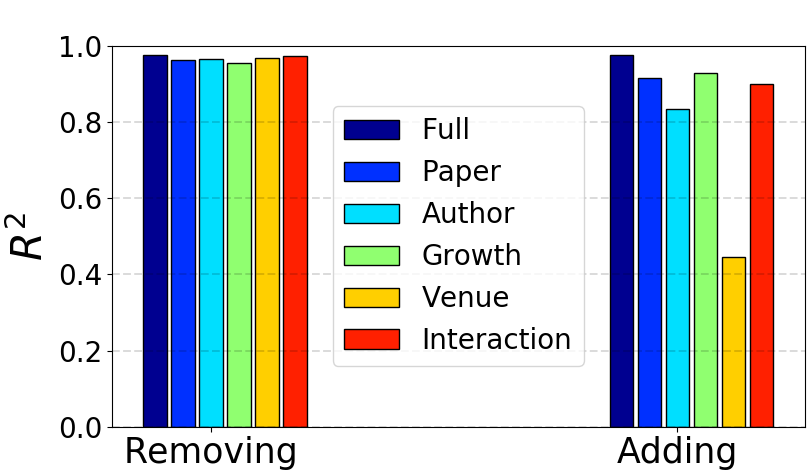}
     \text{\footnotesize{(c)$\Delta t=10$, $R^2$}}
     \label{R2-10-cmp}
   \end{minipage}
   \hfill
   \begin{minipage}[b]{.49\linewidth}
     \centering
     \includegraphics[height=1.0in, width=1.7in]{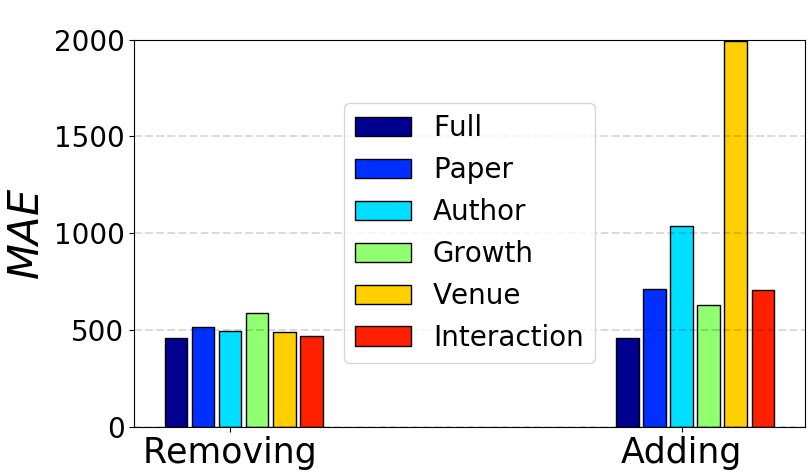}
     \text{\footnotesize{(d)$\Delta t=10$, MAE}}
     \label{MAE-10-cmp}
   \end{minipage}

   \caption{\textbf{Factor contribution analysis.} The performance of prediction using part of factors or only one factor of different prediction interval. "Full" represents the performance of all factors and provides comparison to other factors.}
   \label{fig-cmp}
   \vspace{-5mm}
\end{figure}

To know which part of features in the Table I plays important roles in the prediction, we also measure the prediction performance by "removing" and "adding" factors. As shown in Fig. \ref{fig-cmp}, we illustrate the influences of each factor when the prediction interval $\Delta t=5$ years and $\Delta t=10$ years. The left bars marked by "removing" illustrate the prediction results using factors other than current factor, which means removing the current factor from the training dataset. The right bars marked by "adding" illustrate the prediction results only using the current factor, which equals to add current factor to the empty training dataset.

As we can see that when $\Delta t=5$, the factor of Paper, Author, Growth exhibit good prediction ability and play a dominant role in the whole prediction. This conforms to the result presented in Fig. \ref{fig-features}. In contrast, the remaining factors fail to provide as much information as these major factors. However, the Venue factor and Interaction factor are not totally uninformative useless therefore unnegligible, since both two factors help to improve the total performance cooperating with other main factors. Paper, Author, Growth factors characterize the present general status of a topic, while the Venue and Interaction factors provide more detail information to the prediction. Furthermore, when $\Delta t=10$, the biggest difference is that the Interaction factors' prediction ability shows great improvement, which means that the influences between topics and the driving effects by the related topics are very important in the long-term development. This phenomenon illustrates that in the long term, a topic's development potential is greatly influenced by the growing environment. Which also prove that the factors we have chosen to describe the external relations are valid.

\subsection{Improving Prediction Accuracy}

Besides these ensemble models mentioned before in this section, we also build a multilayer neural network to predict the topic scale after 5 years. And we find that this neural network can better combine all features and get better performance in the prediction.

This neural network is composed of four layers, with the size of each layer being 5, 6, 6, 5. Lbfgs algorithm is chosen as the solver for weight optimization and the activation function of the hidden layer is Relu.

\begin{figure}[!htbp]
\vspace{-2mm}
   \begin{minipage}[b]{.49\linewidth}
     \centering
     \includegraphics[height=1.4in, width=1.6in]{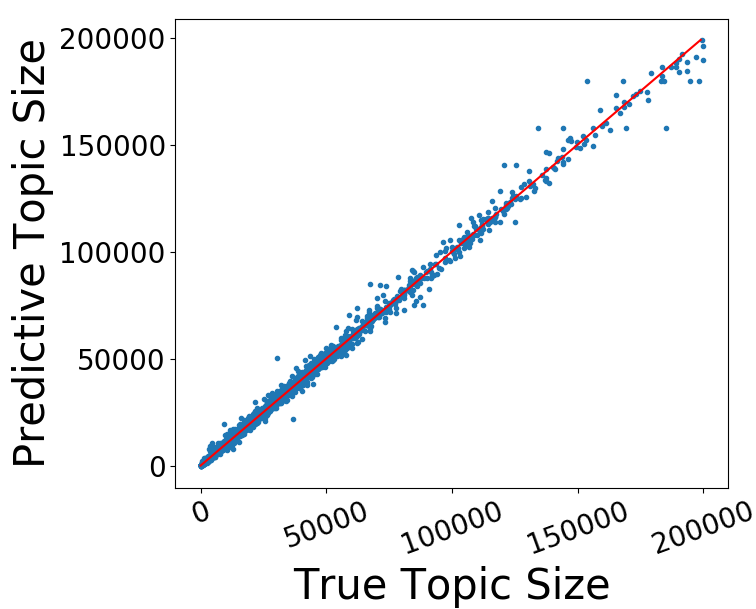}
     \text{\footnotesize{(a) $\Delta t=5$}}
     \label{subfig-1:ann}
   \end{minipage}
   \hfill
   \begin{minipage}[b]{.49\linewidth}
     \centering
     \includegraphics[height=1.4in, width=1.6in]{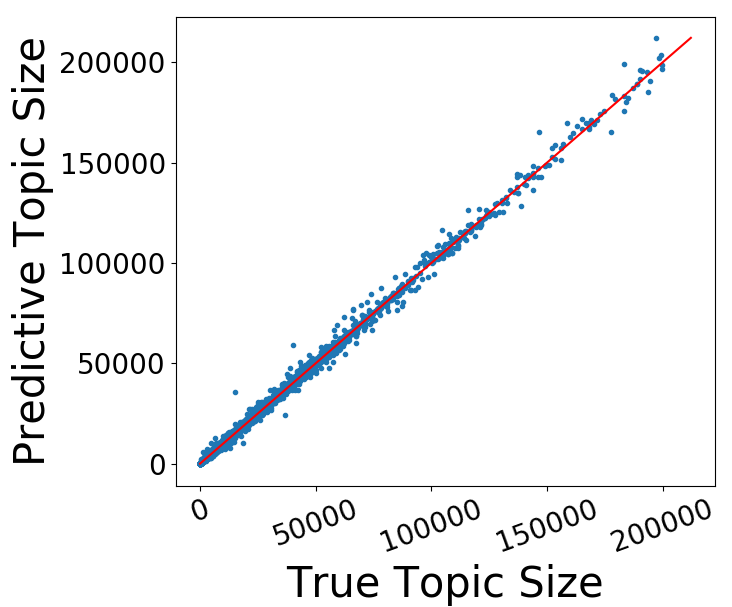}
     \text{\footnotesize{(b) $\Delta t=10$}}
     \label{subfig-2:ann}
   \end{minipage}
   \vspace{-7mm}
   \caption{\textbf{Prediction performance of multilayer neural network.} (a)$\Delta t=5$ years (b)$\Delta t=10$ years.}
   \label{fig:ann}
\vspace{-2mm}
\end{figure}

In our experiment, based on this four layer neural network structure, we get some amazing results. The $R^2$ score computed by this neural network is 0.9983 when $\Delta t=5$ years and 0.9983 when $\Delta t=10$ years, which is relatively high compared to the six ensemble models. As for the MAE score, the neural network model reduces it to 129.32 when $\Delta t=5$ years and 138.88 when $\Delta t=10$ years. The Fig. \ref{fig:ann}. shows the deviation between true values and predictive values of all the test samples which obviously improves the result of ensemble models.

However, our model also bears some limitations that remains to be further improved as our future attention. The best parameters for the model, including layer numbers, nerve connections and etc., are selected by abundant experiments, but it remains unclear what the theoretical mechanism is behind the good performances. The existing work cannot explain the performance of neural network very well and we will research on it in the future.

\subsection{Visualization System: Topic Map}

\begin{figure}[!t]
\centering
\includegraphics[width=3.5in]{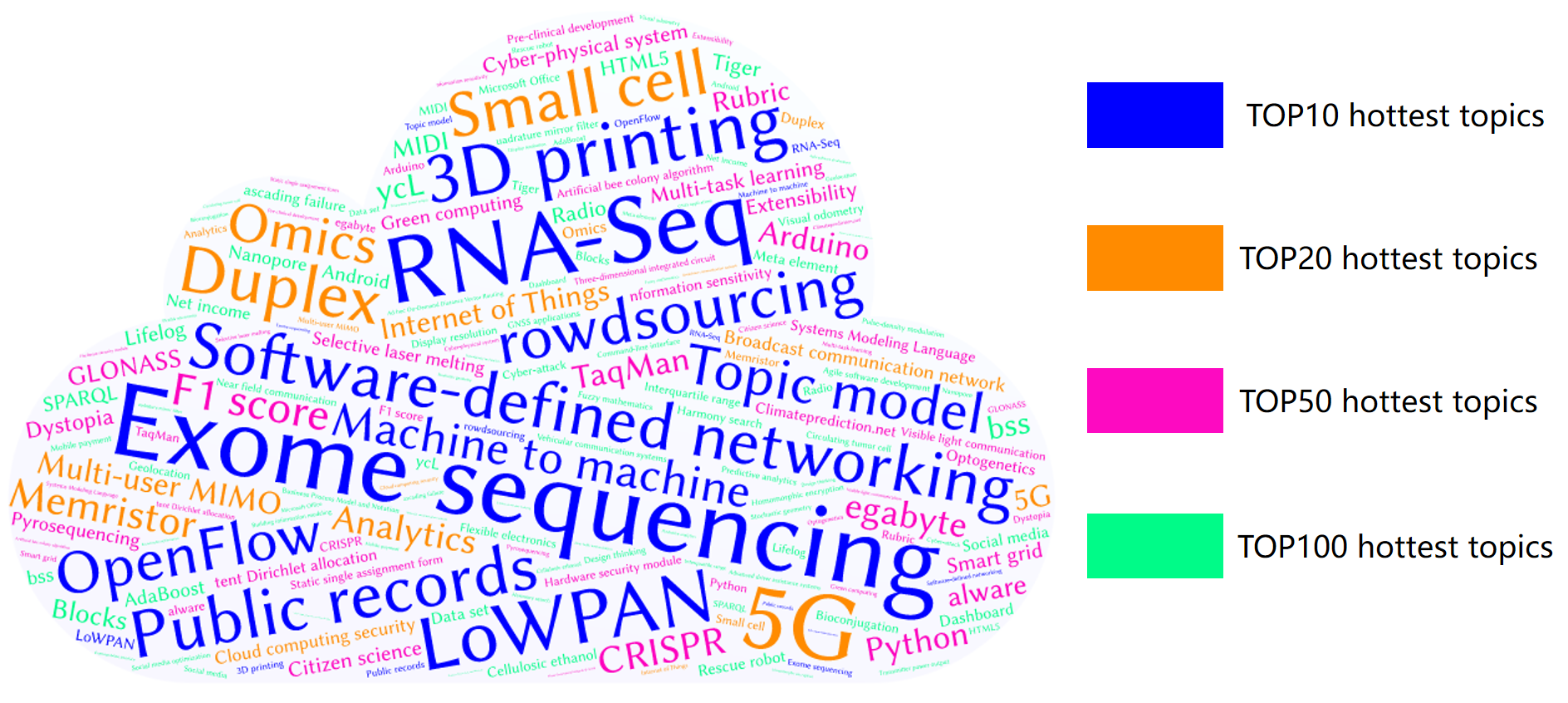}
\vspace{-7mm}
\caption{\textbf{Top 100 hot topics.} The top 100 topics with maximum growth rates in our prediction in the future.}
\label{fig-cloud}
\vspace{-3mm}
\end{figure}

From our prediction, we can determine the future scale of a topic. Based on our dataset, we use the topic information we extracted in 2015 to predict the topic scale 5 years later and find out the top 100 hot topics in the future which have maximum growth rates. From Fig. \ref{fig-cloud}. we can see that some topics like 5g and 3d printing have been hot recently, which prove that our prediction has high accuracy. There is a great possibility that these topics will get more attentions by the researchers and have greater research value and application prospect.

\begin{figure}[!t]
\centering
\includegraphics[width=3.5in]{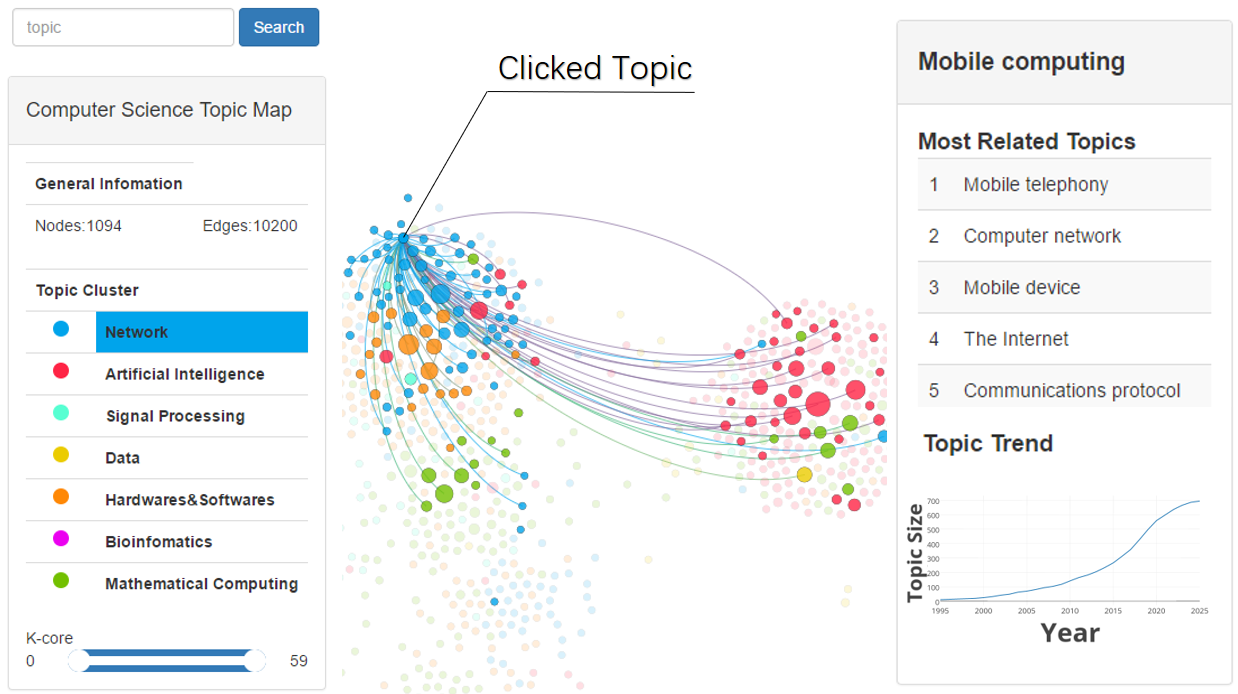}
\vspace{-5mm}
\caption{\textbf{Topic Map.} When click a topic, the Topic Map will show the information of clicked topic, including topic's category, name, and top 5 related topics. Users can also search a specific topic or filter out topics in a specific range of \emph{k-core} value.}
\label{fig-click}
\vspace{-5mm}
\end{figure}

In order to facilitate the usage of our prediction mechanism for users, we deploy an on-line visualization system, named Topic Map, to show the statistical and predictive information of each topic. An overview of the Topic Map is provided in Fig. \ref{fig-click}. Through the system users can search a specific topic they are interested in by the search box, and the topic's position will show on the Topic Map. Using the \emph{k-core} slider at the left, users can assign a range of \emph{k-core} value and filter out topics in the appointed range. Users can also find the basic information of the graph such as the number of edges and node after adjusting the \emph{k-core} slider, which can help users to know the dynamic changes of the graph after \emph{k-core} filtering.

More information inside a certain topic can also be available to users by the click effect that we develop into our Topic Map system. As illustrated in Fig. \ref{fig-click}, when we click one topic on the Topic Map, the select part and related topics of the selected one will be highlighted. The category of the selected topic will be stressed on the left, while the name of the selected topic, along with the top 5 topics that are most correlated with the selected topic will also be listed at the right. From this graph, we can determine the relationship between each topic directly and clearly.

We remark that the goal of the designed Topic Map system can help researchers to gain a clear understanding of the related topics, development history and future trend of topics they major in. We also expect researchers to use this tool to find inspiration from cross domain and make better choices.

\vspace{-1mm}
\section{Related Work}

Traditionally, the topic trend prediction always focusses on the topics they extracted from the texts of small sets of papers. Hurtado $et$ $al.$ \cite{Hurtado2016Topic} extracted topics from a collection of documents and forecasted topic trends. Some work proposed evolving models to predict the topic's future trend. Qian $et$ $al.$\cite{qian2014topic} proposed a model based on the relation of papers in one topic and predicted the core-group's life circle. However, these ways are limited by the quantity of data and the generality is not enough. Due to the rarity of the datasets which contain papers' topic information along with the fact that it is a huge workload to obtain the time series of all the features in scholarly topics, there has been very few prior works that predict academic topical future trend in such a large scale.

Among those efforts that have indeed been made for topic trend prediction, one line of work relies on factor extraction. For example, Qian $et$ $al.$\cite{qian2014topic} analyzed the \emph{k-core} relationship of papers, and Emre $et$ $al.$ \cite{sarigol2014predicting} proposed how centrality in the coauthorship network differs between high impact authors and low impact authors and deploy a classifier to predict the papers' citation. In our work, we extend \emph{k-core} to the topic network analysis and get good result in prediction ability. In the subtask to find meaningful authors in one topic, we use pareto principle to classify the authors. Pareto principle often helps a lot to realize that often a minority of inputs can cause the majority of results.\cite{kiremire2011application,arias2016methodology,wallace2009modeling} We take this method to author classification which makes our prediction more accurate.

To the other scholarly network, a lot of work has been focused on the prediction of the impact of one paper or one author. For example, J. Gehrke $et$ $al.$\cite{gehrke2003overview} concentrated on how to predict the future citation number of a paper according to its present citation. Xiao $et$ $al.$ \cite{xiao2016modeling} proposed a model to predict the individual paper citation count over time. Some work focused on the authors, such as Dong $et$ $al.$\cite{dong2016can} examined the author's h-index in five years and proposed a classifier to distinguish whether a previously (newly) published paper will contribute to the authors' future h-index. They all pay attention to small items of the network and lack of an overview of the whole structure of the field. There is some other work about social network development, such as Saha $et$ $al.$\cite{saha2012learning} Zhu $et$ $al.$\cite{Zhu2014Tracking} and Lin $et$ $al.$\cite{Lin2011The}. They detected emerging topics in the social network and tracked the topics' evolving process. However, the social network have many differences from the scholarly network for its volatility and mutable characteristic. In our work, we propose unique features to describe the development of scholarly topics and utilize these features to predict the future trend accurately.

\section{Conclusion}
\vspace{-1mm}

In this work, we study the future trend of scholarly topics by formalizing two problems that can be reduced to the following questions: scientific factor extraction and topic trend prediction. We proposed to parts of factors that can influence the topics' future trend, which we call External Factor and Internal Factor. We explain the relationship between these factors and future trend of topics. Furthermore, we obtain the time series for recent 50 years of each factor of all 12464 topics.

After using the different models to do the prediction and measure the importance of each factor in predicting the future trend about the topic. We find that the maximum values of each factor provide more information of the future trend and topic's development potential, and the interactions between topics are critical in determining the long-term prediction. The hot topics are probably to drive the related topics hot in the future. Finally, we find out the top 100 topics with the highest growth rates in the future and develop an on-line visualization tool to help users to obtain topics' related information and inspirations by our work.

There is still some future work worth studying. While we conducted our work in the CS field, it is necessary to examine and observe the results in other science fields such as mathematics, biology, literature and so on. The correlations between various elements have more valuable informations, to know how they affect each other, more work is needed.





\vspace{-1mm}
\bibliographystyle{IEEEtran}
\bibliography{sigproc,IEEEabrv}

%




\end{document}